\documentclass[aps,pre,twocolumn,showkeys,showpacs,notitlepage,superscriptaddress]{revtex4-1}
\usepackage{amsmath}
\usepackage{graphicx}
\usepackage{hyperref}
\hypersetup{pdftitle=Phenotypic constraints promote latent versatility and carbon efficiency in metabolic networks, pdfauthor={Marco Bardoscia, Matteo Marsili, Areejit Samal}}

\DeclareMathOperator*{\argmax}{arg\,max}
\begin{document}
\title{Phenotypic constraints promote latent versatility \\and carbon efficiency in metabolic networks}
\author{Marco Bardoscia}
\email{marco.bardoscia@gmail.com}
\affiliation{The Abdus Salam International Centre for Theoretical Physics, Trieste, Italy}
\author{Matteo Marsili}
\email{marsili@ictp.it}
\affiliation{The Abdus Salam International Centre for Theoretical Physics, Trieste, Italy}
\author{Areejit Samal}
\email{asamal@imsc.res.in}
\affiliation{The Abdus Salam International Centre for Theoretical Physics, Trieste, Italy}
\affiliation{The Institute of Mathematical Sciences, Chennai, India}
\begin{abstract}
System-level properties of metabolic networks may be the direct product of natural selection or arise as a by-product of selection on other properties. Here we study the effect of direct selective pressure for growth or viability in particular environments on two properties of metabolic networks: latent versatility to function in additional environments and carbon usage efficiency. Using a Markov Chain Monte Carlo (MCMC) sampling based on Flux Balance Analysis (FBA), we sample from a known biochemical universe random viable metabolic networks that differ in the number of directly constrained environments. We find that the latent versatility of sampled metabolic networks increases with the number of directly constrained environments and with the size of the networks. We then show that the average carbon wastage of sampled metabolic networks across the constrained environments decreases with the number of directly constrained environments and with the size of the networks. Our work expands the growing body of evidence about nonadaptive origins of key functional properties of biological networks.
\end{abstract}
\maketitle

\section{Introduction}

A major focus of systems biology has been the elucidation of design principles of metabolic networks \cite{heinrich1996,jeong2000,wagner2001,stelling2002,tanaka2005,palsson2006,papp2009}. Studies on the structure, dynamics, and function of metabolic networks have revealed several system-level properties such as scale-free topology \cite{jeong2000,wagner2001}, modularity \cite{ravasz2002}, bow-tie architecture \cite{ma2003,csete2004}, flux coupling in linear pathways \cite{burgard2004,samal2006}, robustness \cite{edwards2000}, and alternate metabolic flux states \cite{reed2004}. However, much less is understood about the evolutionary forces behind the establishment of these properties in metabolic networks \cite{wagner2007,papp2009}. Thus, there is considerable interest in uncovering the adaptive properties of metabolic networks which are direct products of natural selection and distinguishing them from nonadaptive properties that arise as by-products of selection on other properties \cite{pal2006chance,wagner2007,papp2009}.

The basic function of a metabolic network is to use available resources in its environment to produce the energy and the precursors required for the growth of an organism. A recent line of research has investigated whether certain system-level properties of metabolic networks could arise as by-products of direct selective pressure of growth in particular environments. One such study \cite{samal2011random} showed that the large-scale structural properties of metabolic networks such as scale-free topology and bow-tie architecture could arise as by-products of biochemical and phenotypic constraints of growth in particular environments. Another study \cite{samal2011env} showed that modularity in metabolic networks could arise as a by-product of phenotypic constraints associated with growth in many different environments. Furthermore, it was shown that a metabolic network subject to direct selective pressure of growth in one particular minimal environment acquires the \textit{latent capacity} to grow in additional minimal environments \cite{samal2010,barve2013}. In this paper, we extend this line of research \cite{pal2006chance,wagner2007,papp2009,samal2011random,samal2011env,samal2010,barve2013} to investigate whether two system-level properties of metabolic networks, latent versatility to function in additional environments and carbon or nitrogen wastage, are affected by the level of direct selective pressure corresponding to the number of minimal environments in which networks are directly constrained to be viable.

The paper is organized as follows: in Sec.\ \ref{methods} we describe the modeling framework to sample random viable metabolic networks that differ in the level of direct selective pressure. In Sec.\ \ref{results} we report and discuss our results. Finally, in Sec.\ \ref{conclusions} we summarize our findings and draw some conclusions.

\section{Modeling framework}
\label{methods}
The metabolism of an organism can be viewed as a collection of processing units, the biochemical reactions, transforming some input, the nutrients, into some other output, energy and biomass. Biochemical reactions form a complex interconnected network in which the output metabolites of some reactions are actually intermediate products constituting the input metabolites of other reactions. From this perspective the metabolism can be visualized as a bipartite network whose two node types are biochemical reactions and metabolites. If a metabolite participates in a reaction, then an edge is present between the corresponding nodes. Every metabolite takes part in a reaction with a given stoichiometric coefficient, which can be thought as a weight associated with the edge. As a consequence, the information about the structure of the network is encoded in the stoichiometric matrix $\mathbf{S}$ of dimensions $m \times n$, where $m$ is the number of metabolites, $n$ is the number of reactions, and $S_{ij}$ is the stoichiometric coefficient of metabolite $i$ in reaction $j$. The stoichiometric matrix also accounts for transport reactions involving import or export of external metabolites across the cell boundary. Denoting by $\mathbf{x}$ the vector of concentrations of metabolites and by $\mathbf{v}$ the vector of reaction fluxes (i.e.\ rates), the rate of change of concentration of metabolites is given by
\begin{equation}
\label{dynamic}
\frac{d \mathbf{x}}{d t} = \mathbf{S} \mathbf{v} \, .
\end{equation}
In addition to the structural properties of the network, thermodynamic and enzyme capacity constraints can affect the reversibility of reactions and limit the flux at which reactions take place, respectively.


\subsection{Flux balance analysis (FBA)}

Flux balance analysis (FBA) is a constraint-based modeling method widely used to analyze the functional capabilities of large-scale metabolic networks \cite{kauffman2003,feist2008,orth2010}. FBA is attractive as it does not require knowledge of metabolite concentrations or detailed information of enzyme kinetics, which are unknown for the majority of metabolic reactions. FBA essentially requires the structural information contained in the stoichiometric matrix $\mathbf{S}$, plus the information about additional constraints on fluxes. FBA assumes a metabolic steady state, such as would be attained by an exponentially growing cell population with adequate nutrient supply, to predict the reaction fluxes and the optimal biomass production rate in a given environment. In any metabolic steady state, from Eq.\ \eqref{dynamic}, the vector $\mathbf{v}$ of reaction fluxes satisfies the equation
\begin{equation}
\label{steadystate}
\mathbf{S} \mathbf{v} = 0 \, .
\end{equation}
Since, the number of metabolites $m$ is typically smaller than the number of reactions $n$ in the metabolic network of an organism, Eq.\ \eqref{steadystate} gives an underdetermined linear system of equations relating various reaction fluxes and leading to a large solution space of allowable reaction fluxes. As already mentioned, the allowable solution space can be further restricted by incorporating known thermodynamic and enzyme capacity constraints associated with certain reactions.

In order to select a specific solution one chooses a biologically relevant quantity to maximize. If such a quantity is a linear function of $\mathbf{v}$, the problem can be solved by means of linear programming. The solution $\mathbf{v}^*$ is then
\begin{equation}
\label{fba}
\mathbf{v}^* = \argmax_{\mathbf{v}} \{ {\mathbf{c}^{T}} \mathbf{v} \, | \,  \mathbf{S}\mathbf{v}=0, \mathbf{a} \leq \mathbf{v} \leq \mathbf{b}\} \, ,
\end{equation}
where $\mathbf{c}$ is the vector of linear coefficients, and vectors $\mathbf{a}$ and $\mathbf{b}$ contain the lower and upper bounds, respectively, of different reaction fluxes. Transport reactions are used to uptake or excrete external nutrient metabolites across the cell boundary. Positive flux of a transport reaction signifies uptake of the external nutrient metabolite, while negative flux signifies excretion of the external nutrient metabolite. In order to include a given metabolite in the environment one has to \emph{enable} the corresponding transport reactions by setting the relative value of $b$ larger than zero. The objective function ${\mathbf{c}^{T}} \mathbf{v}$ is usually taken to be the rate of biomass production. Biomass production is captured through a fictitious reaction that drains biomass precursor metabolites, such as amino acids and nucleotides, in experimentally measured proportions for the growth of the cell. Assigning index $n+1$ to such fictitious reaction one has:
\begin{equation}
c_i = 0 \quad \forall i : 1 \leq i \leq n \, ,
\end{equation}
while $c_{n+1}$ can be incorporated into $v_{n+1}$. Predictions from FBA and related methods are often in good agreement with experimental results \cite{edwards2001,ibarra2002,segre2002}.

\subsection{Biochemical reaction universe}

The Kyoto Encyclopedia of Genes and Genomes (KEGG) LIGAND \cite{goto2002} database represents our present knowledge of the set of biochemical reactions known to occur in some organisms across the three domains of life. In this study, we use a hybrid database compiled by Rodrigues and Wagner \cite{rodrigues2009} containing 4816 metabolites and 5870 reactions, derived mainly from the KEGG LIGAND database \cite{goto2002} after appropriate pruning to exclude mass imbalanced reactions and generalized polymerization reactions. In addition to reactions in the KEGG LIGAND database, this hybrid database also contains the set of reactions in the \textit{E.\ coli} metabolic model iJR904 \cite{reed2003}. More than 90\% of the reactions in the hybrid database are contained in the KEGG LIGAND database and less than 300 reactions are specific to the \textit{E.\ coli} metabolic model iJR904. Of the 5870 reactions in the hybrid database, 2501 are reversible and 3369 are irreversible. Note that the hybrid database also has transport reactions for 143 external metabolites contained in the \textit{E.\ coli} metabolic model iJR904. These 143 external metabolites are assumed to be the set of possible metabolites available for uptake or excretion across the cell boundary. Furthermore, the hybrid database also includes the \textit{E.\ coli} biomass reaction from the metabolic model iJR904 which can be used as the objective function in the FBA framework.

\subsection{Metabolic genotype-to-phenotype map}

Any subset of reactions from the biochemical universe is a possible metabolic network. Following previous work \cite{rodrigues2009,samal2010,samal2011env,samal2011random}, we here take the biochemical universe to be the set of reactions in the hybrid database \cite{rodrigues2009}. Any metabolic network (e.g.\ of \textit{E.\ coli}) is a subset of the global reaction set that can be represented by a binary string of length $N$, i.e.\ $G = (b_1,b_2,\ldots,b_N)$. Each position in the bit string $G$ corresponds to one reaction in the global reaction set with each reaction $i$ being either present ($b_i=1$) or absent ($b_i=0$). One can relate such a string to the genotype, in the sense that it contains the complete information relative to the synthetic organisms that we are sampling. We remark that, although the set of reactions in our global reaction set is most likely incomplete, it is sufficient to produce a vast genotype space of possible metabolic networks with each network containing a subset of reactions. For example, the \textit{E.\ coli} metabolic network contains a subset of $n=931$ reactions. The number of other possible metabolic networks containing exactly $n=931$ reactions forms a vast genotype space of $\approx 10^{1113}$ metabolic networks in the global reaction set.

For any metabolic network (genotype), we are interested in the phenotype defined as the ability to sustain growth in a given environment. Using FBA, we determine whether a metabolic network has the ability to produce all biomass components in a given environment. The phenotype of a metabolic network is considered to be \textit{viable} (respectively, \textit{non-viable}) in a given environment if and only if the maximum biomass production rate for the network predicted by FBA is nonzero (respectively, zero) \cite{samal2010}. We use the \textit{E.\ coli} biomass reaction as the objective function in FBA to determine the viability of a metabolic network in a given environment.

\subsection{Library of minimal environments}
\label{library}

Following previous work \cite{barrett2005,samal2008}, we generate a comprehensive library of different minimal environments containing a carbon, nitrogen, phosphorus, sulfur, and electron-acceptor source as follows. In our global reaction set there are 143 external metabolites that any metabolic network can possibly uptake from the environment. These 143 external metabolites are categorized into possible carbon, nitrogen, phosphorus, sulfur, or electron-acceptor sources. Note that some external metabolites are categorized into multiple categories (e.g.\ glutamate serves as both a carbon and a nitrogen source). Also, each category contains the fictitious element \emph{None}, specifying the unavailability of any metabolite from that category. A complete library of minimal environments is then generated by enumerating all the combinations of metabolites from each category, i.e.\ all the distinct environments containing one carbon source, one nitrogen source, one phosphorous source, one sulfur source and one electron acceptor source, which results in 108,723 possible minimal environments \cite{barrett2005}. Using FBA, we determine for each of the 108,723 minimal environments if a hypothetical \textit{superorganism} equipped with all 5870 reactions in our global reaction set is viable. Of the 108,723 minimal environments, we find that such a hypothetical superorganism is viable in a subset of $V_{\text{super}} =$ 27,646 minimal environments. The subset of $V_{\text{super}}$ minimal environments in which the hypothetical superorganism is viable is then taken to be the library of possible minimal environments. Note that, by construction, any metabolic network that is a subset of the global reaction set cannot be viable in a minimal environment in which the hypothetical superorganism is non-viable.

\subsection{Sampling of random viable metabolic networks}
\label{sampling}

The metabolic networks with fixed number $n$ of reactions from the global reaction set forms a vast genotype space. However, in a previous work \cite{samal2010}, it was estimated that the probability that a random metabolic network with $n=1,000$ reactions from the global reaction set is viable in a single minimal environment is less than $10^{-40}$. This probability decreases even more if one imposes viability in multiple minimal environments. Such tiny probabilities of finding random viable metabolic networks make it computationally infeasible to directly sample viable metabolic networks from the global reaction set. Recently, a new method \cite{samal2010} combining Markov Chain Monte Carlo (MCMC) sampling and FBA was developed to uniformly sample random viable metabolic networks with a fixed number $n$ of reactions and the desired phenotype of viability in specified environment(s). Our study relies on this MCMC-based method to generate large samples of random viable metabolic networks with desired phenotypes from the global reaction set.

The starting point for the MCMC method is an initial metabolic network with $n$ reactions and the desired phenotype of viability in specified environment(s). The method then produces a chain of metabolic networks so that the $(k+1)^{th}$ network in the chain is generated from the $k^{th}$ network using a probabilistic transition rule. Each Markov chain transition step attempts to introduce a small modification in the reaction content of the current metabolic network in the chain. The method then uses FBA to evaluate the phenotype of the modified metabolic network in terms of viability in specified environment(s). If the modified metabolic network has the desired phenotype, one accepts the modified network as the next network of the chain, otherwise the modification is rejected and the next network of the chain is identical to the current one. The modification introduced at each Markov chain step is a reaction swap, which consists of the removal of one reaction from the current metabolic network followed by the addition of a new reaction from the global reaction set to generate a modified metabolic network. The reaction swap preserves the number $n$ of reactions in sampled metabolic networks \cite{samal2010}. Investigation of properties in sampled metabolic networks with a fixed number $n$ of reactions and the desired phenotype avoids artifacts arising due to variable metabolic network size \cite{samal2010}. Note that in the limit of long chains of sampled networks, this method samples uniformly the space of viable metabolic networks with the desired phenotype and connected by reaction swaps starting from the initial metabolic network.

In our MCMC simulations, starting with the initial metabolic network with $n$ reactions and the desired phenotype, we first perform $t_{\text{eq}}$ Markov chain steps (or attempted reaction swaps) to erase the memory of the initial network. Our tests show that $t_{\text{eq}} = 10^5$ is enough for the system to equilibrate. After this preliminary phase, we continue the MCMC to sample metabolic networks with the desired phenotype. However, successive metabolic networks in the chain have nearly identical reaction content (they differ only by two reactions), and so their properties can be expected to be strongly correlated. Hence, one needs to wait a certain decorrelation time before obtaining an independent sample. Thus, we save metabolic networks separated by $t_{\text{dec}}$ Markov chain steps (or attempted reaction swaps). Also in this case the tests we performed showed that essentially $t_{\text{dec}} = 10^3$ is sufficient to make two successive networks uncorrelated for all the considered network sizes.

To begin the MCMC simulation, an initial metabolic network with $n$ reactions and the desired phenotype of viability in specified environment(s) is required. This initial metabolic network is generated as follows. Starting with a network containing all the reactions in the global reaction set, one attempts to remove a sequence of randomly chosen reactions, always checking via FBA after each reaction removal whether the reduced network has the desired phenotype of being viable in the specified environment(s). If so, the reaction removal is accepted, otherwise another reaction is randomly chosen for removal. Hence, we ensure that each randomly chosen reaction removal preserves the desired phenotype of viability in the specified environment(s) for the reduced network. We continue to remove reactions until we reach a reduced network of the desired size of $n$ reactions and use this reduced network as the  initial metabolic network of the MCMC simulation to sample random viable networks.

\section{Results and Discussion}
\label{results}

We define the \emph{environmental versatility} $V_{\text{cons}}$ of sampled metabolic networks in an ensemble as the number of minimal environments in which the sampled networks are directly \emph{constrained} to be viable. Thus, random viable metabolic networks directly constrained to be viable in one specified environment have phenotype $V_{\text{cons}}=1$, in two specified environments they have phenotype $V_{\text{cons}}=2$, and so on. Therefore, considering the members of the sample as synthetic organisms subject to a selection mechanism, environmental versatility is a measure of the direct selective pressure exerted on them.

Using the technique described in Sec.\ \ref{sampling} we build various samples of metabolic networks characterized by different values of number of reactions $n$ and environmental versatility $V_{\text{cons}}$. Moreover, since the properties of the metabolic networks could depend on the specific choice of environments, for a given value of $V_{\text{cons}}$ we resample several times, every time using a different pool of environments in which the metabolic networks are constrained to be viable.

In general, the MCMC sampling is computationally very expensive \cite{samal2010}. As $V_{\text{cons}}$ increases, it becomes even more so \cite{samal2011env} due to two reasons. First, the Monte Carlo acceptance rate becomes smaller as it is more difficult to find a network that is viable in more environments. Second, it becomes computationally expensive to perform FBA for additional environments as $V_{\text{cons}}$ increases to check the viability of networks. A small acceptance rate should be expected also for small values of $n$, precisely because the networks become under-equipped in terms of reaction content. For larger values of $n$ the computational limit is FBA itself, as $n$ is the dimension of the space of the optimization problem defined in Eq.\ \eqref{fba}.

Since \textit{E.\ coli} is a generalist \cite{reed2003} that can thrive in many environments, it makes good sense to compare the properties of random viable metabolic networks against the functional capabilities of the \textit{E.\ coli} metabolic network iJR904, which has $n = 931$ reactions. Taking the aforementioned limits into account we then choose $n = $ 400, 700, 931, and 1200 and $V_{\text{cons}} = 1, 2, \ldots, 10$. For every pair of values $(n, V_{\text{cons}})$ we generate 10 samples (with each sample corresponding to a different pool of $V_{\text{cons}}$ environments) of 500 networks each. In addition, for $n = 931$ we also consider $V_{\text{cons}} = 20, 40, \ldots, 200$ and generate 5 samples (with each sample corresponding to a different pool of $V_{\text{cons}}$ environments) of 200 networks each.

\subsection{Latent versatility}

The first quantity we focus on is the \textit{latent versatility} $V_{\text{latent}}$ of a random metabolic network, defined as the total number of minimal environments in which the network is viable. In our setting $V_{\text{latent}}$ is a simple proxy that quantifies how many metabolic features are acquired by a synthetic organism as a by-product of direct selective pressure. To compute $V_{\text{latent}}$ for a given network we use FBA to check its viability across the $V_{\text{super}}$ minimal environments in our library.

\begin{figure}
\includegraphics[width=.85\columnwidth]{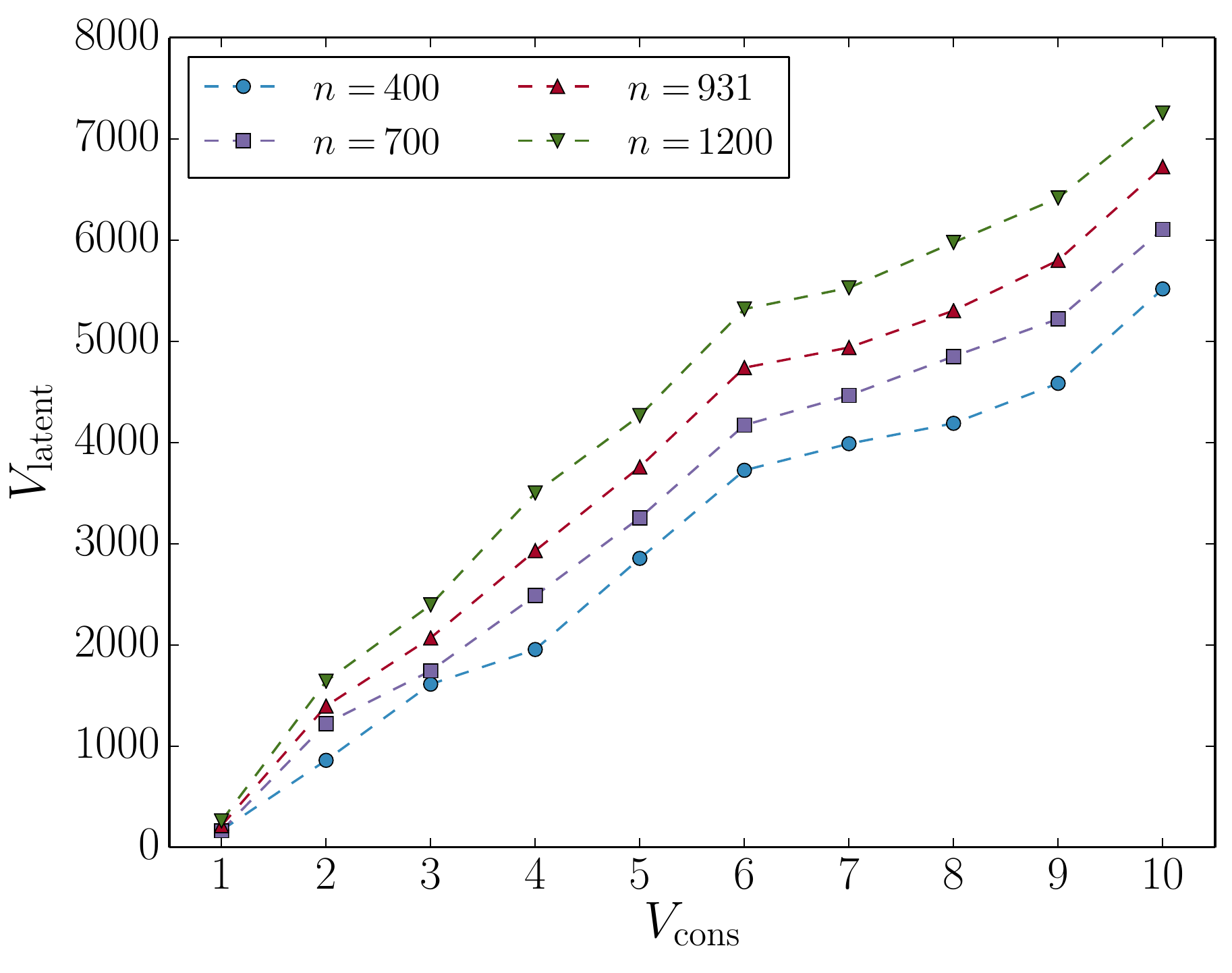}
\caption{Environmental versatility $V_{\text{cons}}$ is shown on the horizontal axis and latent versatility $V_{\text{latent}}$ of random viable metabolic networks is shown on the vertical axis. Each data point is obtained by averaging over 10 different pools of $V_{\text{cons}}$ minimal environments with each pool corresponding to a sample of 500 random viable metabolic networks with a fixed number $n$ of reactions. Error bars are within the symbol size.}
\label{latent}
\end{figure}

In Fig.\ \ref{latent} we plot $V_{\text{latent}}$ as a function of $V_{\text{cons}}$, for different values of $n$. First, let us note that, already for $V_{\text{cons}} = 1$, we find $V_{\text{latent}} \approx 10^2$, for all the values of $n$. Up to $V_{\text{cons}} = 6$, latent versatility increases roughly linearly, so that by constraining for viability in one additional environment (i.e.\ $V_{\text{cons}} \rightarrow V_{\text{cons}} + 1$) one gains the capability to be viable in about 700 additional environments for $n = 400$, about 800 additional environments for $n = 700$, about 900 additional environments for $n = 931$, and about 1,000 additional environments for $n = 1200$. A direct selective pressure for viability in only 10 minimal environments implies a functional capability to be viable in at least 5,000 minimal environments, for all network sizes. As already noted in Sec.\ \ref{library}, there is an upper bound for $V_{\text{latent}}$ given by the number of environments in which a hypothetical superorganism is viable. Hence, at some point, $V_{\text{latent}}$ must start to saturate and become concave. From Fig.\ \ref{latent} we see that for $V_{\text{cons}} > 6$, $V_{\text{latent}}$ already seems to deviate from linearity. As expected, $V_{\text{latent}}$ also increases with $n$, as networks equipped with more metabolic reactions are viable in a larger set of environments.

\begin{figure}
\includegraphics[width=.85\columnwidth]{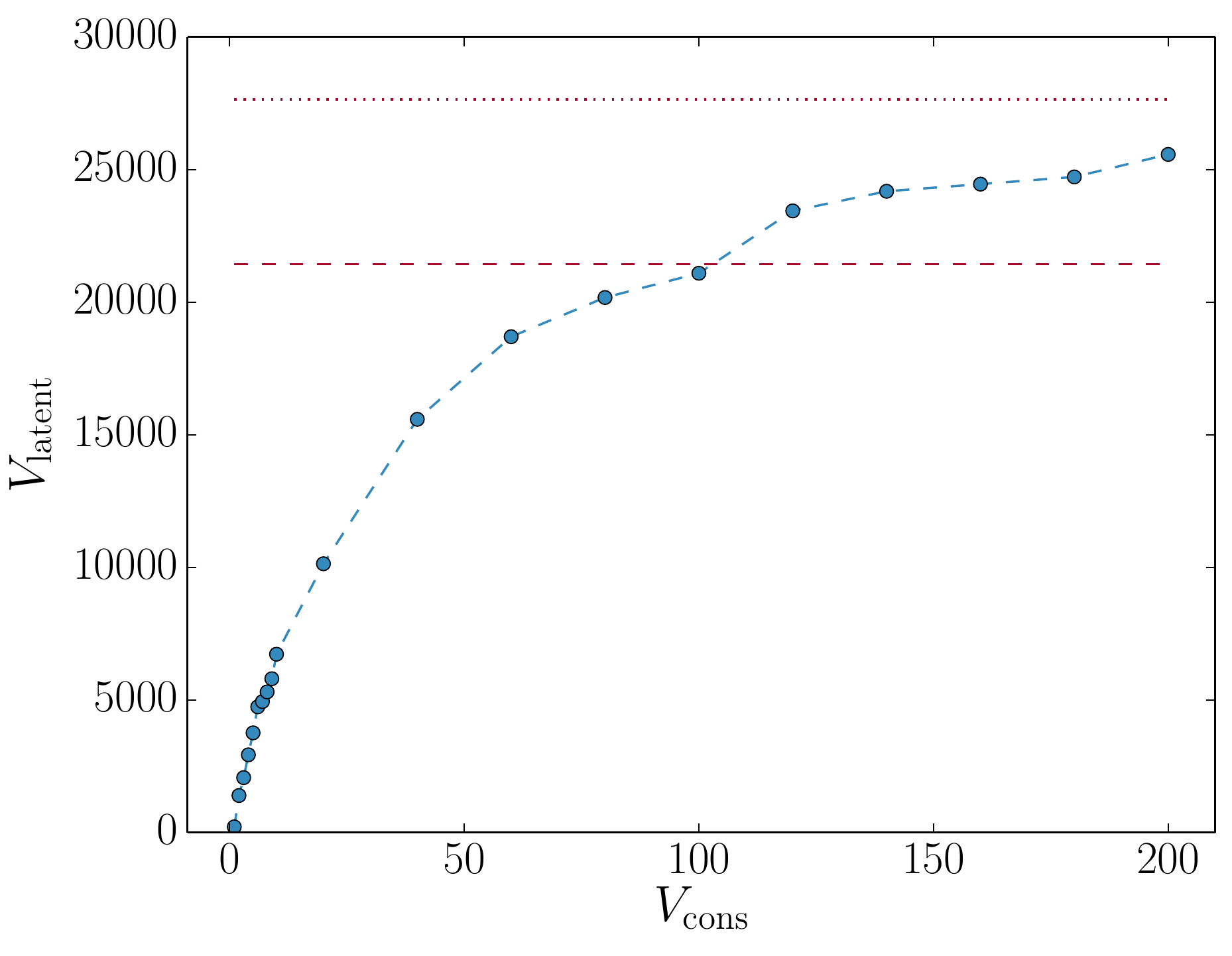}
\caption{Environmental versatility $V_{\text{cons}}$ is shown on the horizontal axis and latent versatility $V_{\text{latent}}$ of random viable metabolic networks with $n=931$ (as in the \textit{E.\ coli} metabolic network iJR904) is shown on the vertical axis. Each data point is obtained by averaging over 10 (5) different pools of $V_{\text{cons}}$ minimal environments with each pool corresponding to a sample of 500 (200) random viable metabolic networks for $V_{\text{cons}} \leq 10$ ($V_{\text{cons}} \geq 20$). Error bars are within the symbol size. The dashed line corresponds to $V_{\mathit{E.\,coli}}$, the number of minimal environments in which the \textit{E.\ coli} metabolic network iJR904 is viable. The dotted line corresponds to $V_{\text{super}}$, the total number of minimal environments in which a hypothetical superorganism equipped with all reactions in our global reaction set is viable.}
\label{latentecoli}
\end{figure}

The \textit{E.\ coli} metabolic network iJR904 contains a subset of reactions from our global reaction set (for details see Sec.\ \ref{methods}). Using FBA, we check the viability of the \textit{E.\ coli} metabolic network in each of the $V_{\text{super}}$ possible minimal environments in our library, finding that \textit{E.\ coli} is viable in $V_{\mathit{E.\,coli}} = $ 21,434 minimal environments. However, from Fig.\ \ref{latent}, one can see that $V_{\text{latent}} \approx $ 6000 for random networks with $n =$ 931 reactions (i.e.\ the same number of reactions as in the \textit{E.\ coli} metabolic network iJR904) and subject to direct selective pressure $V_{\text{cons}}$ of viability in 10 minimal environments. Thus, we next address the following question: what is the level of direct selective pressure $V_{\text{cons}}$  required to render $V_{\text{latent}}$ for random networks of the same size as \textit{E.\ coli} equal to $V_{\mathit{E.\,coli}}$? In Fig.\ \ref{latentecoli} we show that, if $V_{\text{cons}} \approx $ 100, then $V_{\text{latent}}$ is equal to $V_{\mathit{E.\,coli}}$, for random networks with $n =$ 931 reactions. In our setting this result implies that by exerting a direct selective pressure for viability in about 100 minimal environments on a random metabolic network equipped with the same number of reactions as in \textit{E.\ coli}, one is actually able to recover (on average) the same functional capability as that of \textit{E.\ coli} itself.

\subsection{Carbon efficiency}

For any metabolic network, FBA finds an optimal flux distribution that maximizes the biomass production given the available supply of external nutrient metabolites in the considered environment (for details see Sec.\ \ref{methods}). Although the optimal flux distribution for a metabolic network achieves maximum biomass yield, typically not all of the carbon, nitrogen, and other elements imported in the form of external nutrient metabolites is converted into biomass. Thus, the optimal flux distribution for a metabolic network also leads to some waste of carbon, nitrogen, or other elements in the form of excreted external metabolites \cite{schuster2008,molenaar2009}.

We define the \textit{carbon wastage} of a metabolic network in a given environment, based on the optimal flux distribution predicted by FBA, as the fraction of carbon excreted as waste per unit of carbon intake. Note that  \textit{carbon usage efficiency} or \textit{carbon yield} of a metabolic network in a given environment is simply given by 1 minus carbon wastage. The \textit{nitrogen wastage} and \textit{nitrogen efficiency} can be analogously defined. The amount of carbon (nitrogen) intake per external metabolite is computed as the product of the number of carbon (nitrogen) atoms in an external metabolite and the uptake flux of that external metabolite in the optimal flux distribution predicted by FBA. Similarly, the amount of carbon (nitrogen) excretion per external metabolite is computed as the product of the number of carbon (nitrogen) atoms in an external metabolite and the excretion flux of that external metabolite in the optimal flux distribution predicted by FBA. To obtain the total amount of carbon excreted as waste and the total amount of carbon intake, one has to sum over all the external metabolites.

In computing such quantities an important detail is that in FBA a given environment is specified by setting the upper bounds and the lower bounds of transport reactions associated with the uptake of available external metabolites and, usually, one of the available external metabolites in the environment for intake is the limiting source \cite{orth2010}. Hence, one could in principle compute the optimal flux distribution by limiting either the carbon source or the nitrogen source. To check that the obtained results are not an artifact of FBA, we computed the carbon wastage and the nitrogen wastage using both schemes: carbon as the limiting source and nitrogen as the limiting source.

\begin{figure}
\includegraphics[width=.85\columnwidth]{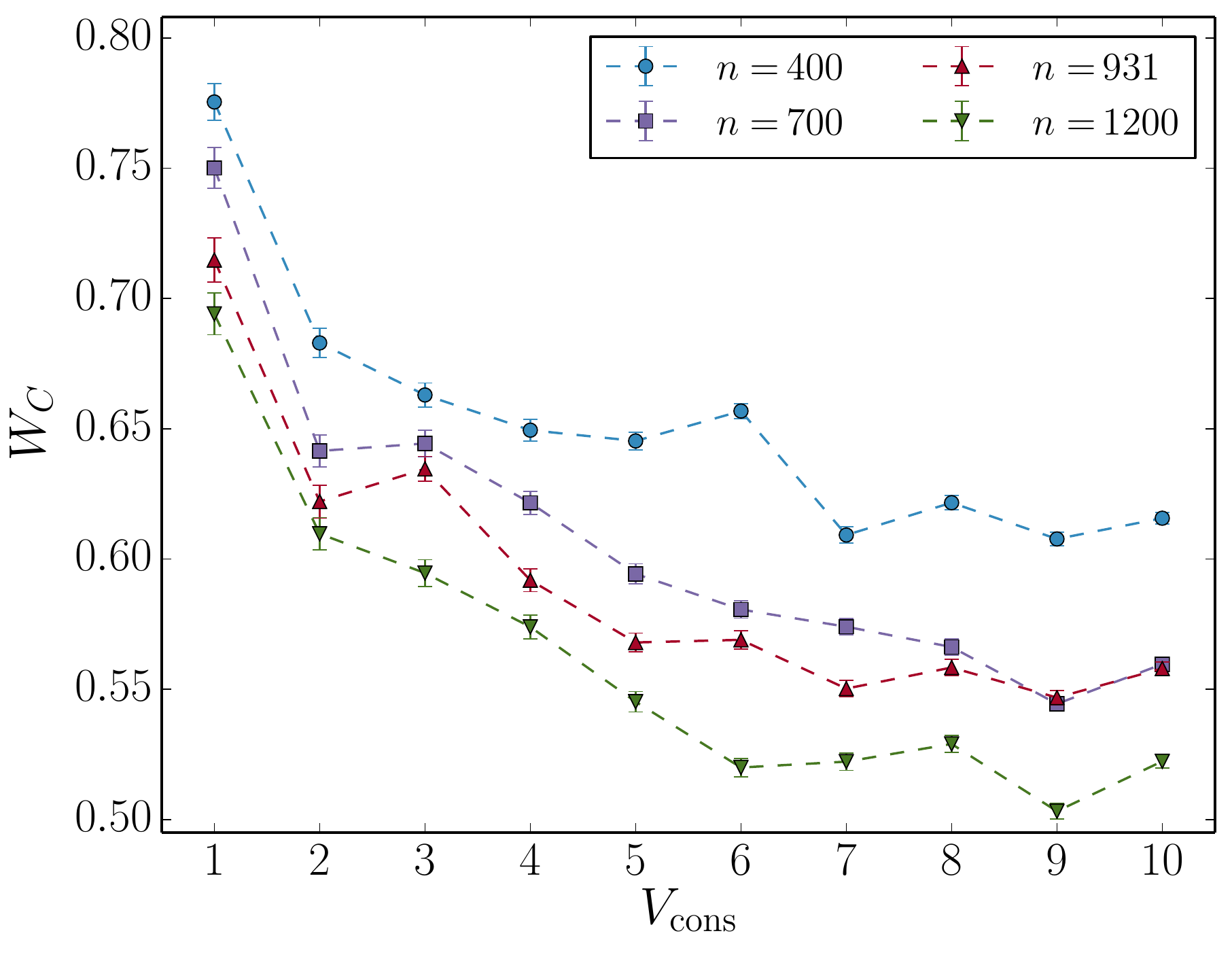}
\caption{Environmental versatility $V_{\text{cons}}$ is shown on the horizontal axis and average carbon wastage $W_C$ across the pool of $V_{\text{cons}}$ minimal environments in which the networks are constrained to be viable is shown on the vertical axis. Each data point is obtained by averaging over 10 different pools of $V_{\text{cons}}$ minimal environments with each pool corresponding to a sample of 500 random viable metabolic networks with a fixed number $n$ of reactions. Error bars are within the symbol size.}
\label{carbonwaste}
\end{figure}

For a metabolic network sampled with the technique presented in Sec.\ \ref{sampling}, i.e.\ of fixed size $n$ and constrained to be viable on a given pool of $V_{\text{cons}}$ environments, we define $W_C$ ($W_N$) as the carbon (nitrogen) wastage averaged over all the environments in which the metabolic network is constrained to be viable. In Fig.\ \ref{carbonwaste} we plot $W_C$ (computed with carbon as the limiting source) as a function of $V_{\text{cons}}$. From Fig.\ \ref{carbonwaste} one can see that the average carbon wastage decreases as the number of minimal environments in which the metabolic networks are constrained to be viable increases. Such a behavior is certainly nontrivial, if not counterintuitive. In fact one might expect that an organism constrained to be viable on less minimal environments is more specialized and therefore able to exploit the resources at its disposal more efficiently. For carbon wastage it appears that the opposite is true: metabolic networks subject to a selective pressure that makes them viable in more minimal environments are also able to use carbon more efficiently (in the environments they have been selected for). From Fig.\ \ref{carbonwaste} we also show that $W_C$ decreases as $n$ increases. Such an effect might be the consequence of the fact that metabolic networks equipped with more reactions have a larger choice of metabolic pathways, some of which will be more efficient than the others. Results obtained for nitrogen as the limiting source are undistinguishable from those presented in Fig.\ \ref{carbonwaste}.

\begin{figure}
\includegraphics[width=.85\columnwidth]{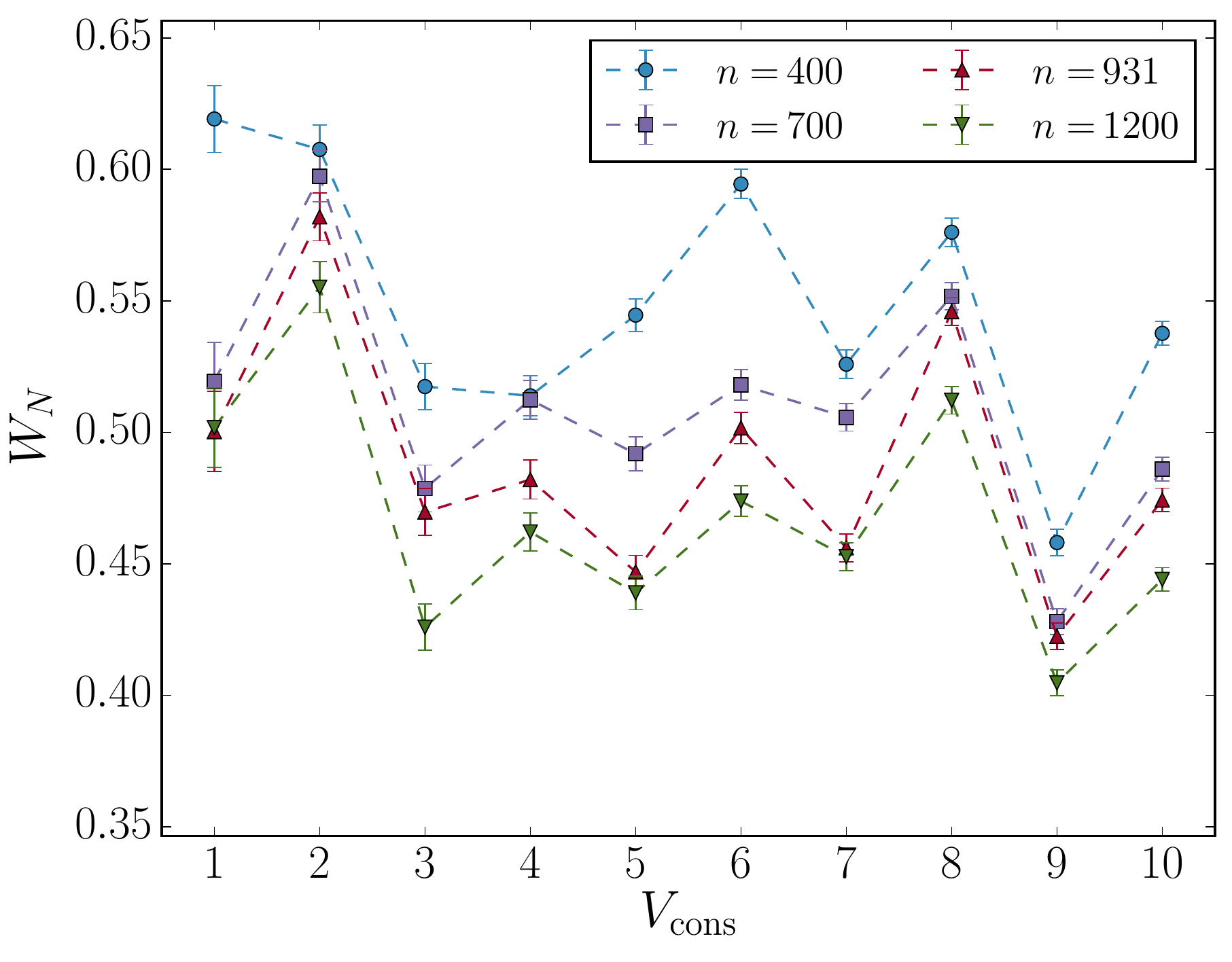}
\caption{Environmental versatility $V_{\text{cons}}$ is shown on the horizontal axis and average nitrogen wastage $W_N$ across the pool of $V_{\text{cons}}$ minimal environments in which the networks are constrained to be viable is shown on the vertical axis. Averages were computed as in Fig.\ \ref{carbonwaste}.}
\label{nitrogenwaste}
\end{figure}

In Fig.\ \ref{nitrogenwaste} we plot the average nitrogen wastage $W_N$ (computed with nitrogen as the limiting source) as a function of $V_{\text{cons}}$. Quite remarkably, $W_N$ is uncorrelated with $V_{\text{cons}}$, thus behaving in a completely different way than $W_C$. However, comparably to $W_C$ in Fig.\ \ref{carbonwaste}, for a given value of $V_{\text{cons}}$, $W_N$ decreases as the size $n$ of metabolic networks increases. Results obtained for carbon as the limiting source are undistinguishable from those presented in Fig.\ \ref{nitrogenwaste}.

\begin{figure}
\includegraphics[width=.85\columnwidth]{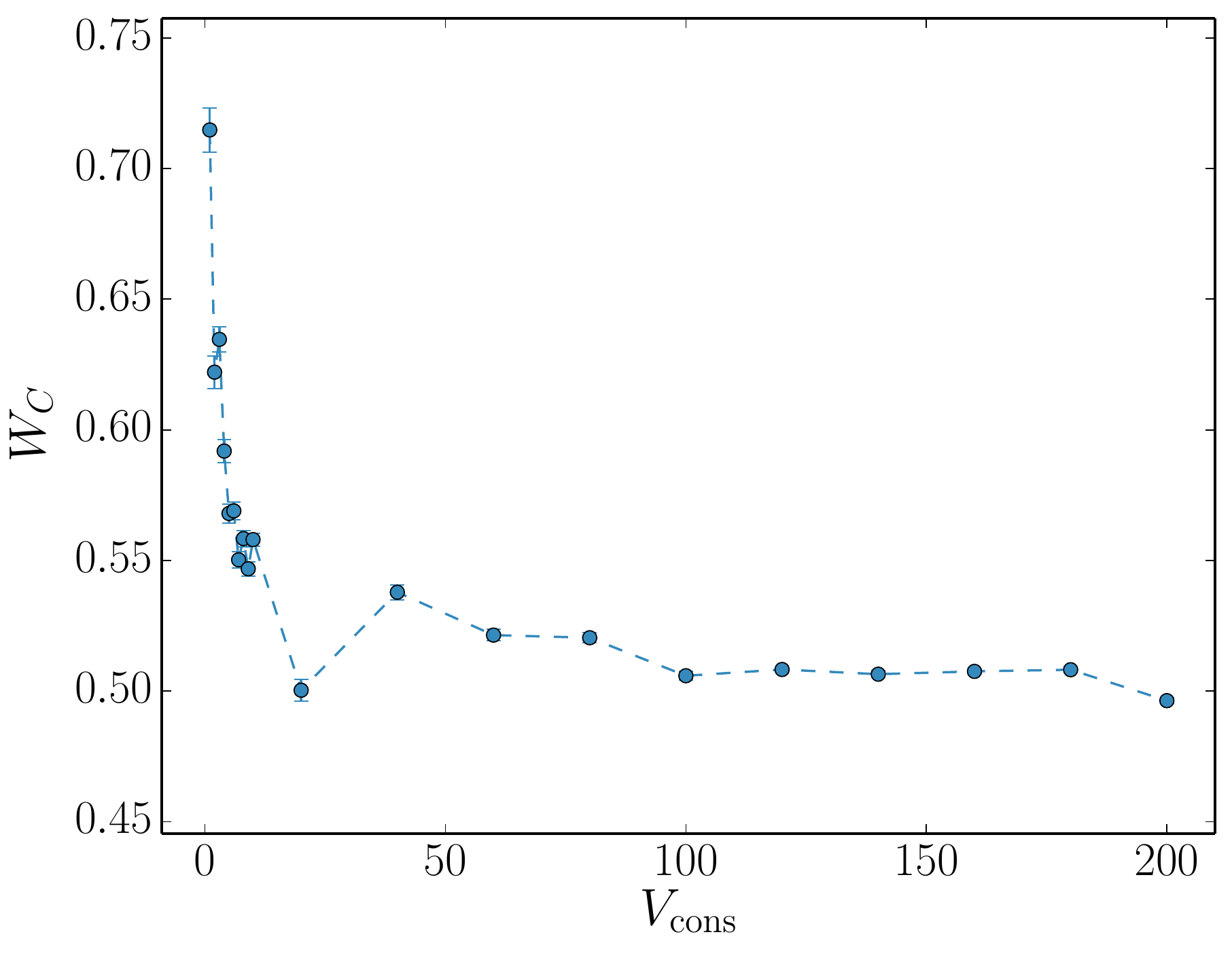}
\caption{Environmental versatility $V_{\text{cons}}$ is shown on the horizontal axis and average carbon wastage $W_C$ across the pool of $V_{\text{cons}}$ minimal environments in which networks with $n = 931$ are constrained to be viable is shown on the vertical axis. Each data point is obtained by averaging over 10 (5) different pools of $V_{\text{cons}}$ minimal environments with each pool corresponding to a sample of 500 (200) random viable metabolic networks for $V_{\text{cons}} \leq 10$ ($V_{\text{cons}} > 10$). Error bars are within the symbol size.}
\label{carbonecoli}
\end{figure}

\begin{figure}
\includegraphics[width=.85\columnwidth]{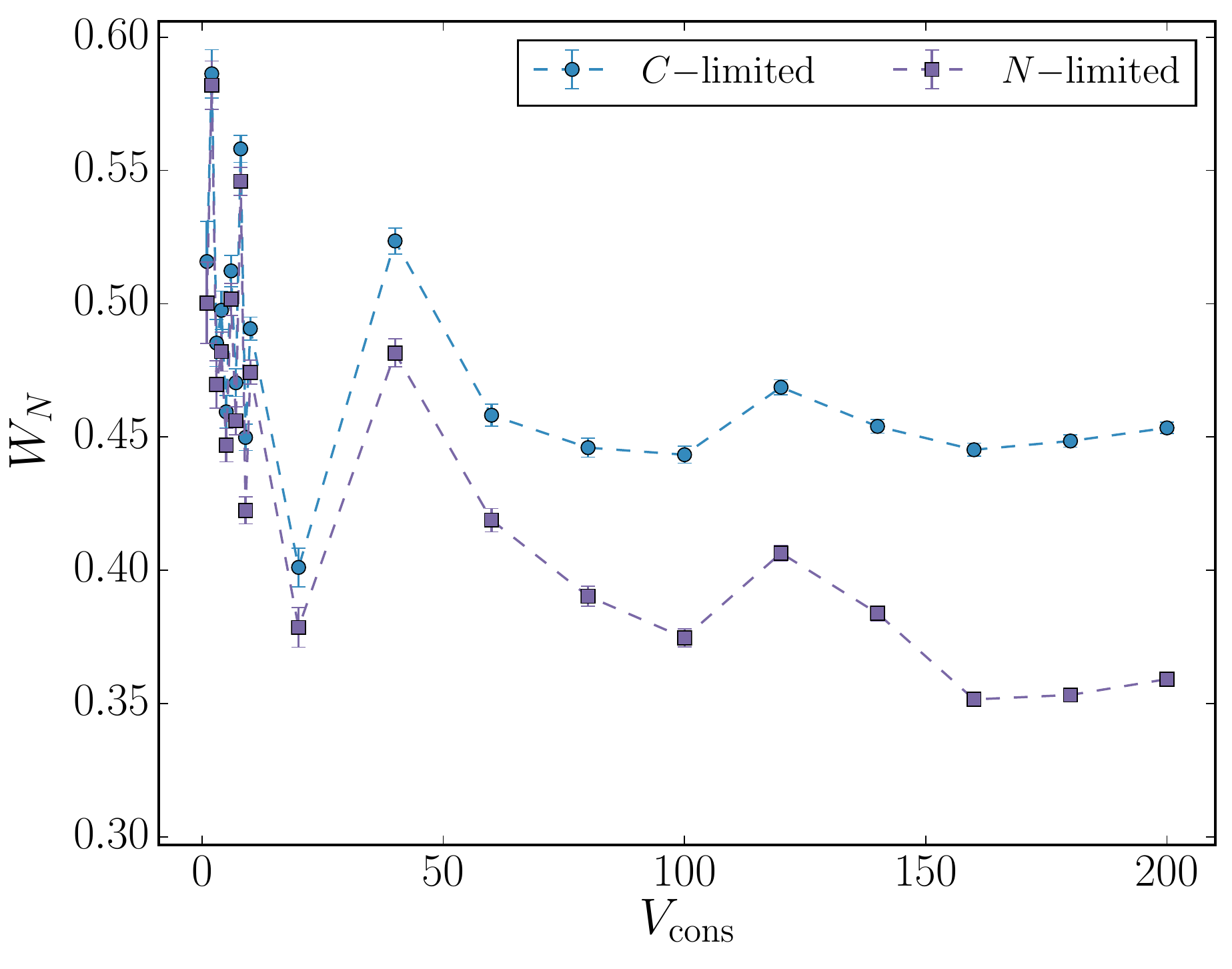}
\caption{Environmental versatility $V_{\text{cons}}$ is shown on the horizontal axis and average nitrogen wastage $W_N$ across the pool of $V_{\text{cons}}$ minimal environments in which networks with $n = 931$ are constrained to be viable is shown on the vertical axis for both carbon and nitrogen as limiting sources. Averages were computed as in Fig.\ \ref{carbonecoli}.}
\label{nitrogenecoli}
\end{figure}

As for latent versatility, for networks with the same number of reactions as in the \textit{E.\ coli} metabolic network iJR904, we explore a broader range of values of $V_{\text{cons}}$. In Fig.\ \ref{carbonecoli} we show that, interestingly, $W_C$ (computed with carbon as the limiting source) reaches a plateaux for $V_{\text{cons}}$, meaning that the carbon usage efficiency does not improve indefinitely as $V_{\text{cons}}$ increases. Results obtained for nitrogen as the limiting source are undistinguishable from those presented in Fig.\ \ref{carbonecoli}. As regards $W_N$, and as shown in Fig.\ \ref{nitrogenecoli}, the trend is not as clear as for $W_C$, for the two cases of nitrogen or carbon as the limiting source.

These findings suggest that carbon usage efficiency, which (as already mentioned) is equal to $1 - W_C$, is increasing with the level of direct selective pressure $V_{\text{cons}}$. As regards the nitrogen usage efficiency, our results in Figs.\ \ref{nitrogenwaste} and \ref{nitrogenecoli} do not show the presence of such a consistent trend.


\subsection{Pathway level analysis}

We next studied the pathway level differences in the reaction content of the sampled metabolic networks to shed light on the possible underlying biological reasons for the increase in latent versatility and carbon efficiency with the selective pressure. The idea is to look at the reactions that \emph{occur} in the sampled networks belonging to an ensemble with a large value of $V_{\text{cons}}$, but do \emph{not occur} in networks belonging to an ensemble with a smaller value of $V_{\text{cons}}$. We fixed $n =$ 931 reactions (i.e.\ the same number of reactions as in \emph{E.\ coli}) and, to filter some noise, we considered only reactions appearing in at least half of the networks in the two ensembles. Given the computational time constraints, we were able to sample 5000 networks (10 different pools of $V_{\text{cons}}$ minimal environments and 500 networks per pool) only for $V_{\text{cons}} \leq 10$, and so it was natural to choose $V_{\text{cons}} = 10$ as the larger value of $V_{\text{cons}}$ in this analysis. Now, in order to identify a set of reactions distinguishing networks in the ensemble with $V_{\text{cons}} = 10$ from those in an ensemble with a smaller value of $V_{\text{cons}}$, it is necessary to choose $V_{\text{cons}}$ as small as possible. In fact, in the ensemble with $V_{\text{cons}} = 1$ we found 221 reactions that occur in more than half of the sample, while in the ensemble with $V_{\text{cons}} = 10$ we found 198 reactions that occur in more than half of the sample, resulting in 40 reactions that occur in the second sample, but not in the first one.

\begin{figure}
\includegraphics[width=.85\columnwidth]{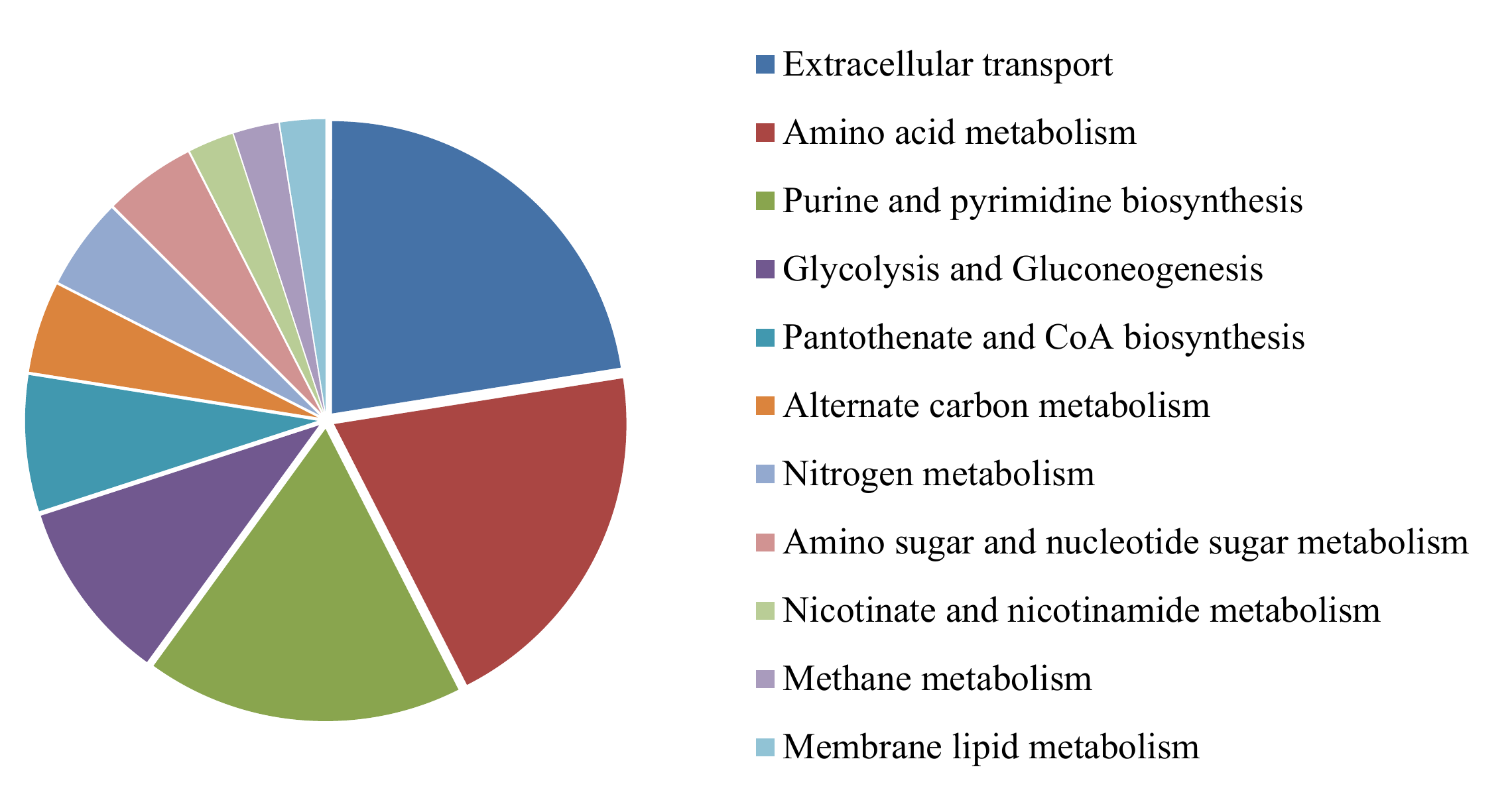}
\caption{Pie chart depicting the enrichment of KEGG metabolic pathways in the set of reactions that occur in more than half of the sampled networks with $n =$ 931 reactions and $V_{\text{cons}} = 10$, but do not occur in more than half of the sampled networks with $n =$ 931 reactions and $V_{\text{cons}} = 1$.}
\label{pathwayenrichment}
\end{figure}

We studied the enrichment of different KEGG \cite{goto2002} metabolic pathways in this set of 40 reactions to identify specific pathways responsible for the increase in latent versatility and carbon efficiency with the number of directly constrained environments. From Fig.\ \ref{pathwayenrichment} it can be seen that most of these 40 reactions belong to four metabolic pathways: extracellular transport, amino acid metabolism, purine and pyrimidine metabolism, and glycolysis and gluconeogenesis. Since sampled networks with $V_{\text{cons}} = 1$ are directly constrained to be viable in fewer environments compared to sampled networks with $V_{\text{cons}} = 10$, it is natural that the later sample has many more transport reactions to uptake nutrients in additional constrained environments. Furthermore, observed differences at the level of amino acid metabolism, purine and pyrimidine metabolism, and glycolysis and gluconeogenesis suggest that sampled networks with $n =$ 931 reactions and $V_{\text{cons}} = 10$ possibly have additional reactions to efficiently generate proteins, nucleic acids, and carbohydrates from available nutrients, when compared to sampled networks with $n =$ 931 reactions and $V_{\text{cons}} = 1$.

\section{Conclusions}
\label{conclusions}

In this paper we investigate whether two system-level properties of metabolic networks, latent versatility and carbon usage efficiency, are affected by the level of direct selective pressure and show that such properties can arise as by-products of phenotypic constraints of viability in many different environments. Using a novel method based on MCMC and FBA, we are able to sample a large number of metabolic networks with a fixed number of reactions that are constrained to be viable in a given pool of minimal environments.

We find that latent versatility of sampled metabolic networks, measured as the total number of minimal environments in which the networks are viable, increases with the number of minimal environments in which metabolic networks are directly constrained to be viable and also increases with the number of reactions in the networks. Although previous works \cite{samal2010,barve2013} have shown that a sampled metabolic network directly constrained to be viable in one particular minimal environment acquires the latent versatility to grow in additional minimal environments, we explore in detail here the dependence of latent versatility of sampled metabolic networks on the number of minimal environments in which sampled networks are directly constrained to be viable. Furthermore, we also show that one is able to recover the functional capability of \textit{E.\ coli} in sampled metabolic networks with the same number of reactions as in \textit{E.\ coli} when the sampled networks are constrained to be viabile in only 100 specified minimal environments. In the future, it will be interesting to extend our analysis to infer the set of minimal environments in which \textit{E.\ coli} is most likely to grow under direct selection pressure.

We next show that the average carbon wastage of random metabolic networks across the $V_{\text{cons}}$ minimal environments in which the networks are constrained to be viable decreases with increases in $V_{\text{cons}}$ and also with increases in the number of reactions in the networks. We again highlight that this result is in contrast to the conventional wisdom that a ``jack of all trades is master of none''. In fact random networks that are more versatile in terms of the number of environments in which they are viable are also more efficient in their carbon usage. Interestingly, such an effect is peculiar to carbon usage efficiency, while a clear trend is not present for nitrogen usage. A possible explanation for this observed difference in the trends of carbon usage efficiency versus nitrogen usage efficiency may lie in the existence of different catabolic pathways within the biochemical universe for generating energy and biomass precursor metabolites from carbon and nitrogen sources.

We conclude with a suggestion for another possible extension of the present work. The present MCMC sampling method \cite{samal2010} uses FBA with maximization of biomass production as the cellular objective to determine the viability of a random metabolic network in a given minimal environment. Using this method it has been shown that two random metabolic networks constrained to be viable in a given minimal environment can differ in more than 70\% of their reaction content for networks with the same number of reactions as in \textit{E.\ coli} \cite{samal2010}, and this result implies the presence of several alternate metabolic routes within the biochemical reaction universe. Although the same biomass yield can be achieved by two random metabolic networks with very different reaction contents, it is possible that the two networks differ also  in terms of the enzymatic cost required to produce the same amount of biomass. In the present MCMC sampling method \cite{samal2010}, simple FBA can be replaced by \emph{flux minimization} \cite{holzhutter2004} or \emph{parsimonious FBA} \cite{lewis2010} to account for enzymatic costs of biomass production. Such a modified MCMC sampling method will be able to sample random viable metabolic networks with the additional constraint of minimizing enzymatic or protein costs, but requiring much more computational time. In the future, it will be interesting to extend the present work to study the possible role of phenotypic constraints corresponding to both viability in minimal environments and minimization of enzymatic costs associated with biomass production on the latent capacity of innovation in metabolic networks.

\begin{acknowledgments}
We thank Antonio Celani, Christopher J.\ Marx, Sergei Maslov, and Michele Vendruscolo for fruitful discussions. We also thank Andreas Wagner and Joao F. Matias Rodrigues for sharing their database of reactions used in this study. Computer programs used to sample viable metabolic networks in this manuscript are available upon request from the authors.
\end{acknowledgments}

\bibliography{Efficiency}

\end{document}